\documentclass[11pt,twocolumn]{article}
\usepackage[T1]{fontenc}
\usepackage[utf8]{inputenc}
\usepackage[a4paper,margin=1in]{geometry}
\usepackage{graphicx}
\usepackage{tabularx}
\usepackage{url}
\usepackage{hyperref}
\usepackage{booktabs}
\usepackage{siunitx}
\usepackage{setspace}
\usepackage{xparse}
\usepackage{caption}
\usepackage{subfig}

\newcommand{\EOD}{}

\newenvironment{slimabstract}
  {\small\setstretch{0.95}}
  {}

\def\figwidth{0.99\columnwidth}

\NewDocumentCommand{\Figure}{ d() O{} m m }{%
  \begin{figure}[t]
    \centering
    \IfNoValueTF{#2}
      {\includegraphics[width=\linewidth]{#3}}
      {\includegraphics[#2]{#3}}
    \caption{#4}
  \end{figure}
}

\def\BibTeX{{\rm B\kern-.05em{\sc i\kern-.025em b}\kern-.08em
    T\kern-.1667em\lower.7ex\hbox{E}\kern-.125emX}}

\begin{document}

\twocolumn[
\begin{center}
{\LARGE\bfseries
SAGE: Tool-Augmented LLM Task Solving Strategies in Scalable Multi-Agent Environments
\par}

\vspace{2em}

{\large
\textbf{Robert K. Strehlow}$^{1,3*}$, 
\textbf{Tobias Küster}$^{1,3}$,  
\textbf{Oskar F. Kupke}$^{1,3}$,
\textbf{Brandon Llanque Kurps}$^{1,3}$,
\textbf{Fikret Sivrikaya}$^{1,3}$, 
\textbf{Sahin Albayrak}$^{1,2}$ 
\par}

\vspace{0.5em}

{\small
$^{1}$Technische Universität Berlin (e-mail: first.second@tu-berlin.de) \\
$^{2}$DAI-Labor, TU Berlin, Chair of Agent Technologies (e-mail: first.second@dai-labor.de) \\
$^{3}$GT-ARC gGmbH (e-mail: first.second@gt-arc.com) \\
\par}

\vspace{0.5em}

{\small
*Corresponding author
\par}
\end{center}

\vspace{1.0em}

{\footnotesize
This paper is based on the research conducted in the GoKI project, funded by the German Federal Ministry of Labour and Social Affairs (BMAS) under the funding reference number DKI.00.00032.21. Furthermore, we acknowledge support by the German Research Foundation and the Open Access Publication Fund of TU Berlin.
\par}

\vspace{3.5em}
]

\begin{slimabstract}
\textbf{Abstract —}
Large language models (LLMs) have proven to work well in question-answering scenarios, but real-world applications often require access to tools for live information or actuation. For this, LLMs can be extended with tools, which are often defined in advance, also allowing for some fine-tuning for specific use cases. However, rapidly evolving software landscapes and individual services require the constant development and integration of new tools.\\
Domain- or company-specific tools can greatly elevate the usefulness of an LLM, but such custom tools can be problematic to integrate, or the LLM may fail to reliably understand and use them. For this, we need strategies to define new tools and integrate them into the LLM dynamically, as well as robust and scalable zero-shot prompting methods that can make use of those tools in an efficient manner.\\
In this paper, we present ``SAGE'', a specialized conversational AI interface, based on the OPACA framework for tool discovery and execution. The integration with OPACA makes it easy to add new tools or services for the LLM to use, while SAGE itself presents rich extensibility and modularity. This not only provides the ability to seamlessly switch between different models (e.g. GPT, LLAMA), but also to add and select prompting methods, involving various setups of differently prompted agents for selecting and executing tools and evaluating the results.\\
We implemented a number of task-solving strategies, making use of agentic concepts and prompting methods in various degrees of complexity, and evaluated those against a comprehensive set of benchmark services. The results are promising and highlight the distinct strengths and weaknesses of different task-solving strategies. Both SAGE and the OPACA framework, as well as the different benchmark services and results, are available as Open Source/Open Data on GitHub.

\end{slimabstract}

\setcounter{footnote}{0}
\renewcommand{\thefootnote}{\arabic{footnote}}


\section{Introduction}
\label{sec:intro}

Although large-language models (LLMs) have been proven to work reliably in a question-answer scenario~\cite{ouyang2022training}, many real-world applications require the usage of live knowledge and telemetry. By extending LLMs with microservices in the form of tools, responses can be elevated by selected on-demand information. Tools are defined in advance and passed to the model upon inference, giving the LLM the choice to formulate one or multiple tool calls as a response instead of a natural language output. Specifically trained models, such as ToolLLaMA~\cite{qin2923toolllm}, which supports 16000+ real-world APIs from RapidAPI, can leverage even multi-stage approaches to fulfill certain requests. However, with rapidly evolving software environments, growing knowledge bases, and the call for individualized service architecture, strategies to include and integrate additional tools have to be developed and able to be used in an efficient way.

Especially domain-specific applications of tools that are developed on-site or are embedded in a company-internal infrastructure, for example mailing services, proprietary knowledge bases, or interactive office devices, can increase the usability, as well as the receptivity of LLM systems in modern work environments. The transformation of such services into LLM-tools should be made understandable and accessible, so a wide range of users are able to develop their own LLM-tools, promoting further integration of diverse services and knowledge basis into an LLM-pipeline alike. By lowering the requirements of AI tool development, an increase of AI technology in multiple industry fields could be realized as well. 

In this work, we present a system that autonomously chooses and calls tools to fulfill user requests. The key aspect of our system is its ability to generalize and dynamically access and call available tools from an independent runtime platform, exposing tools as agent services across a unified API. In addition, the system is able to use different models in zero-shot inference calls, accommodating a wide variety of application fields without overfitting a specific area. Relying on a uniform model integration, the deployment and usage of local models can also be supported through the vLLM framework~\cite{kwon2023efficient}. This approach directly enhances the development process and the availability of new LLM-tools in versatile service environments.

Furthermore, we present different task-solving strategies, consisting of a variety of specialized LLM inferences, dividing the response generation process into smaller subtasks. Finally, we test our system and the implemented task-solving strategies with our own created benchmark set and tool evaluation method, focusing on cross-domain response generation. All of our code base\footnote{\url{https://github.com/GT-ARC/opaca-llm-ui}}, as well as the benchmark containers\footnote{\url{https://github.com/RobertStrehlow/opaca-llm-benchmark-containers}}, are publicly available on GitHub.

The remainder of this paper is structured as follows: We start by giving an overview of the background (Section~\ref{sec:background}) and related work (Section~\ref{sec:related}). Afterwards, we describe the core concepts (Section~\ref{sec:concept}) and architecture (Section~\ref{sec:arch}) of our framework. In Section~\ref{sec:methods} we describe different prompting methods implemented in the system and present a systematic evaluation of these methods, as well as the used LLMs, in Section~\ref{sec:eval}, before wrapping up in Section~\ref{sec:conclusion}.


\section{Background}
\label{sec:background}

\subsection{The OPACA Multi-Agent Framework}
\label{sec:opaca}

An integral part of this work is the multi-agent framework OPACA (Open, Language- and Platform-Independent API for Containerized Agents)~\cite{acar2024opaca}, which manages and orchestrates agents with modern agent-, container-, and microservice technologies. It consists of two main components:

\begin{itemize}
    \item \emph{Agent Containers} (AC), encapsulating one or more Agents, each providing different Actions, which can be invoked by an external event or through proactive, agentic behavior, and
    \item \emph{Runtime Platform} (RP), the core of the framework, managing one or more Agent Containers through Docker or Kubernetes, connecting with other runtime platforms and providing a user management system and token-based authentication.
\end{itemize}

Both components provide a REST API, used for communication between the components and with external services and users. The available routes provide multiple services for interaction with the deployed containers and their agents and to retrieve general information about the platform. An agent container typically holds one or more agents, providing a multitude of different services, or actions, which can be called via the RP onto which the AC has been deployed.

The OPACA framework makes it easy to develop and deploy new applications on a running system and further provides seamless interaction with other applications on that or connected RPs. Applications running in OPACA are self-descriptive, following common microservice guidelines. Supporting both the OpenAPI standard and a more streamlined proprietary format, each AC describes its agents, those agents' actions, their parameters and required data types, as well as optional natural-language descriptions of their purposes, using a combination of OpenAPI, JSON Schema, and OPACA's own JSON-based models. These aspects make the OPACA framework well suited for dynamic, distributed, and heterogeneous systems and facilitate access to those systems using an LLM.

\subsection{Agent Container Development}

The core of the OPACA framework, the Runtime Platform, is publicly available on GitHub\footnote{OPACA on GitHub: \url{https://github.com/GT-ARC/opaca-core}}, while the Agent Containers are intended to be developed by users and later added to a running RP. A reference AC is implemented with Kotlin using JIAC VI~\cite{rakow2019jiacviits}, but ACs can be developed in any programming language, as long as they provide the API specified by OPACA. In addition, a Python SDK has been developed, streamlining the development process for OPACA Agent Containers for advanced and novel users alike by providing high-level functions and decorators that allow to expose any function as an agent action or data stream.

Compared to similar protocols such as the Model Context Protocol (MCP)~\cite{hou2025model}, the OPACA protocol extends the action/tool exposure with an agentic approach, binding tools to specific agents and containers. Later, a centralized interaction point is handling all interaction with available actions, in which a connected user does not need to be concerned with added, removed, updated, or deprecated agent containers, since the platform manager can easily modify the platform during its runtime.

Due to its language-agnostic architecture and the HTTP protocol used in the OPACA framework, further AC implementations in other programming languages are possible and have already been realized. This could further allow for a faster adaption of the OPACA framework across multiple domains with different coding requirements. 


\section{Related Work}
\label{sec:related}

Following the rapid adoption of AI technology in recent years, numerous advances have been made to enhance the general usability and integration possibility of large language models. The basis for this were realized by the increased model performances, achieved by each new model generation of various model families, such as the GPT series of OpenAI, Llama by Meta, Mistral, Gemini and others (cf.~\cite{naveed2024llmOverview} for a recent overview).

Especially the composition of function calls have been explored early on to extend the capabilities of LLMs beyond simple text generation tasks. One of such an approach is Toolformer~\cite{schick2023toolformer}, which has fine-tuned a pre-trained GPT-J model with numerous exemplary API calls. The model was then able to formulate new API calls for novel user inputs in a format that can easily be translated to concrete function calls, simplifying the downstream process of autonomously calling external functions. Other well-known systems, such as ReAct~\cite{yao2023react} enhanced such an approach by introducing reasoning steps between function calls, enabling the model to interpret function call results during intermediate steps.

Function calling has since become an integral part of LLMs, broadening the scope of possible integrations in real-world applications or as an embedded software component for tool orchestration. Hence, the call for reliable benchmark sets measuring the effectiveness for models to formulate function calls has become more necessary. Sets such as CallNavi~\cite{song2025callnavi} especially deal with nested function calls, in which the parameters first need to be retrieved by an additional function call, which presents a challenge even for modern LLM to deal with hidden dependencies. Other approaches, such as API-Bank~\cite{li2023api}, emphasize the different steps, being \textit{call}, \textit{retrieval}, and \textit{planning}, each with the possibility of being evaluated by a tool-augmented LLM. 

Advances have also been made to autonomously generate LLM conform function calls. Approaches such as ToolMaker~\cite{wolflein2025llm} use LLM agents to generate function calls from open-source code bases. Reevaluating its own decision, the function call generation flow in ToolReflection~\cite{polyakov2025toolreflection} reflects its own choice of API calls by using self-generated data, retrieved by the model's and function call's error response.

Furthermore, the growing popularity of the Model Context Protocol (MCP)~\cite{hou2025model} closely resembles the core concept of the OPACA framework itself, although it focuses primarily on function calling for LLMs. The fast and wide adoption of it further showcases the strong need for a reliable function calling architecture, which is easily extendable and customizable. While MCP is mainly used in combination with LLMs, the OPACA framework allows the creation of a more agentic environment, in which tools are mere agent actions and an LLM integration is complementary to the overall design structure. In this way, agents can still communicate with each other, leveraging their microservice architecture advantages, enhancing the built structure with an overarching LLM interaction.


\section{Concept}
\label{sec:concept}

Building upon the foundation laid out in the OPACA framework, we construct and develop an LLM-application interacting with the different ACs of OPACA via its RP. This interaction is simplified by the standardized API, allowing for quick service retrieval and invocation. Further, it allows for the dynamic modification of services available to the LLM at any point, making the integration of this system into a new environment easy and straightforward. This opens up the possibility of the integration of heterogeneous services into the toolbox of an LLM.

\subsection{Design Goals}
\label{sec:sysreq}


In this and the next section we present our developed system ``SAGE'', also known as ``OPACA-LLM'', whose prime functionality is to answer and fulfill user requests with the help of tools, provided as agent actions via the multi-agent framework OPACA. We define the following design goals:

\begin{itemize}
    \item \textbf{Extensibility}: The system can be easily extended with new (alternative or experimental) prompting methods, which are composed of different LLM modules, with different granularity and responsibilities.
    \item \textbf{Model-Independence}: The system and the different methods implemented within it can be used with a variety of different LLM, such as GPT, LLAMA, Mistral, etc., by employing proxies such as vLLM or LiteLLM.
    \item \textbf{Automatic Tool Calling}: After the used LLM generates the formal tool call, our system should automatically call the chosen tool(s), receive the results from the connected OPACA RP, and feed them back to the LLM.
    \item \textbf{Zero-Shot Prompting}: The models used within the system should not be fine-tuned to accommodate a specific use-case. Tools are instead passed to the models at runtime, assuring the most up-to-date environment.
    \item \textbf{Self-Evaluating}: The results of the tool calls should be evaluated by the LLM in the context of the current user request. Follow-up calls can be initiated with additional knowledge gained from the previous iteration.
    \item \textbf{Resource-Efficiency}: The tokens (and hence cost) and time it takes to generate the final result should be analyzed and evaluated. Both the time and the tokens used for LLM calls should be minimized.
\end{itemize}

Note that due to the ambiguous meaning of the term ``agent'', in the following we will use the term `agent' only for the agents within the OPACA system, and the term `module' for what otherwise might be called an ``LLM agent''. Likewise, functions provided by the OPACA agents are referred to as `actions', but we will use the common term `tool' once they have been transferred to the LLM world. A `method' consists of one or more specifically prompted `modules' (i.e., LLM agents) working together.

\subsection{Tool Call Generation Process}
\label{sec:toolgenprocess}

The general process of a request handled by SAGE using tools can be divided into smaller subtasks. This provides a first overview of the identified steps and a possible role division of different modules within our system.

\begin{itemize}
    \item \textbf{Planning}: The model should choose the most fitting tools from the available list of tools, and generate a rough plan what steps have to be taken to reach the goal.
    \item \textbf{Construction}: Given the information from the planning phase, the LLM constructs a well-structured tool call, including the tool name and tool parameters using the correct data types.
    \item \textbf{Evaluation}: The results of the invoked tool calls are analyzed against the initial user request to determine whether the retrieved information can be used to fulfill the request or additional tool calls are required.
    \item \textbf{Output Generation}: All requested information as well as the retrieval process are then summarized in the final response shown to the user.
\end{itemize}

In addition to the above steps, performed by the LLM, the following steps are executed by the surrounding system before and during/after the LLM calls, respectively:

\begin{itemize}
    \item \textbf{Acquisition}: Retrieve and prepare agent actions from the connected OPACA RP into a well-defined tool-format.
    \item \textbf{Validation}: Iterate over the constructed tool parameters and validate their data types against the data types given within the OPACA agent action definitions, try to perform a simple type casting if necessary.
    \item \textbf{Calling}: Given the information from the selection phase, the tool/action gets called via the standardized API of the OPACA RP. The results are returned in a well-defined format to the LLM.
\end{itemize}


\section{Architecture}
\label{sec:arch}

\Figure()[width=\figwidth]{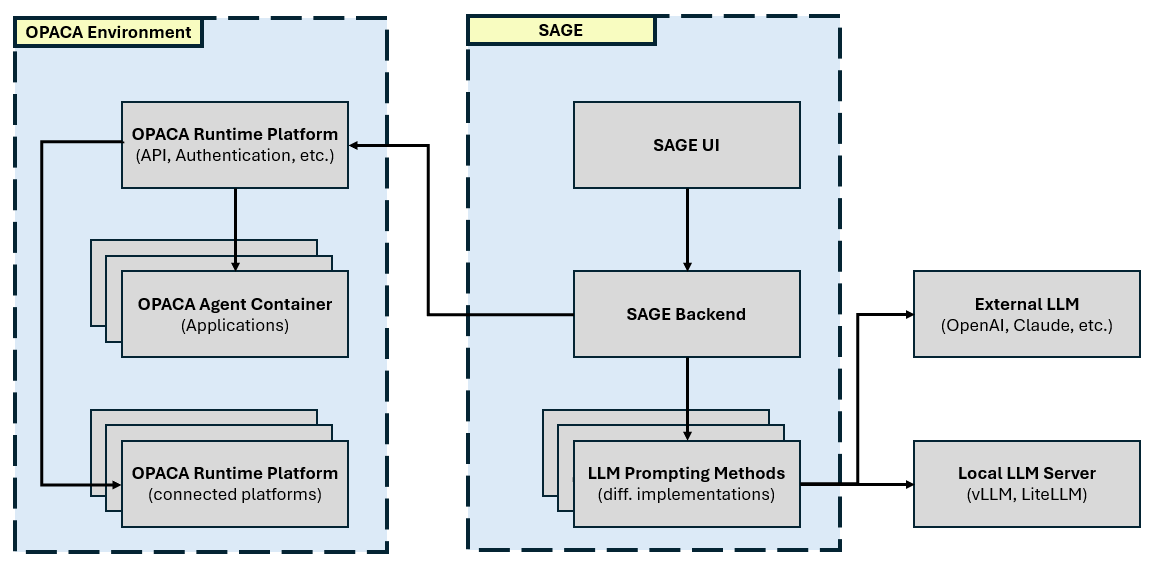}{SAGE System Architecture\label{fig:opaca-llm-arch}}

\Figure()[width=\figwidth]{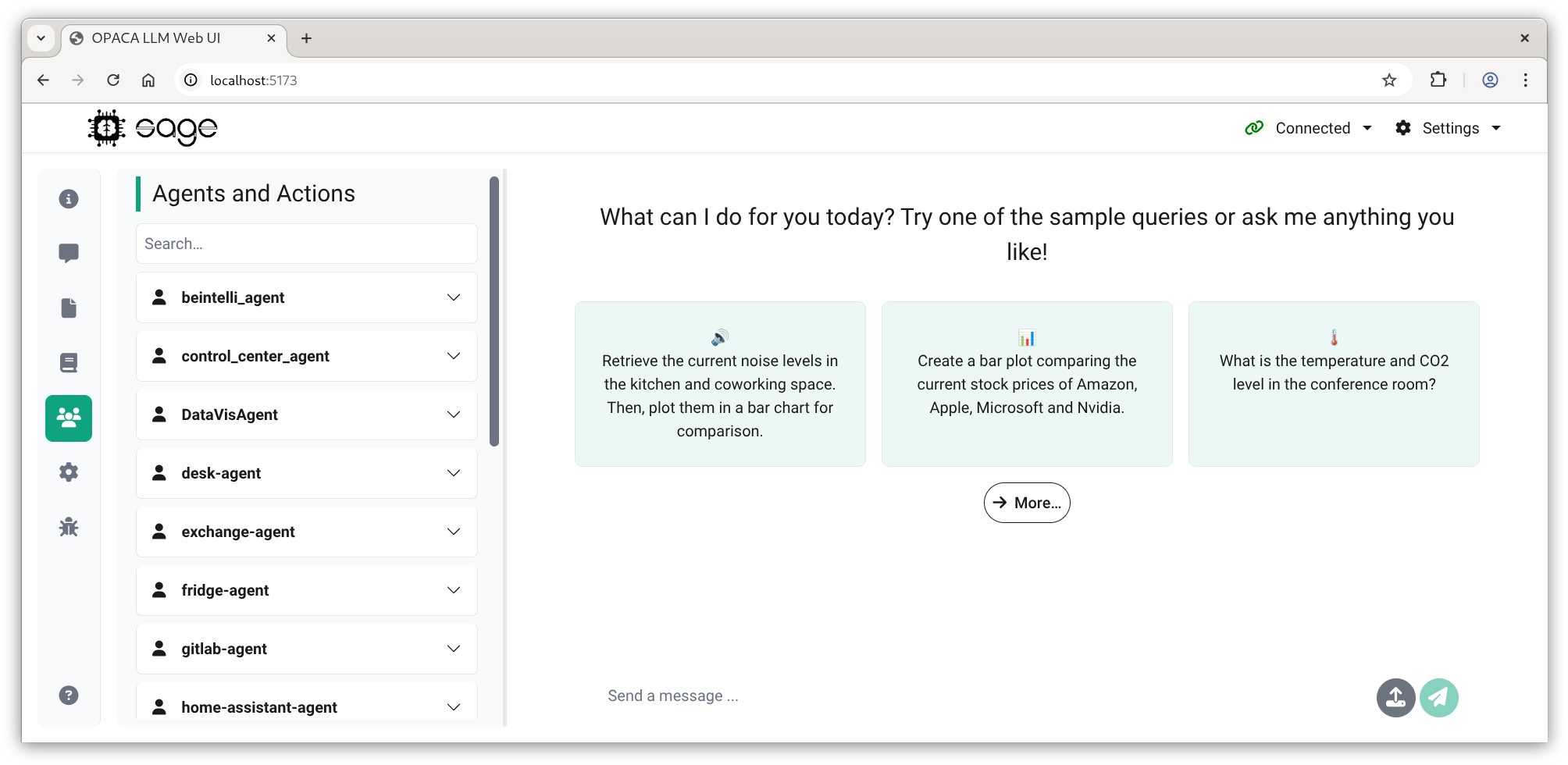}{Screenshot of SAGE, showing `agents' panel and empty chat window with example prompts.\label{fig:opaca-llm-ui}}

The system architecture of SAGE consists of two main components:

\begin{itemize}
    \item \emph{Frontend}, developed with JavaScript using the Vue framework to create interactive UI features, and
    \item \emph{Backend}, developed with Python using FastAPI and providing a REST API to interact with its functionality.
\end{itemize}

Additionally, a running OPACA RP is required, to which the backend connects upon request. While for the core functionality, only the backend is required to run, the frontend simplifies the interaction with SAGE immensely by providing a clean and user-friendly interface with an overview of different functionalities, which are used to send HTTP requests to the backend. In the backend, these requests are handled by a REST API; an additional websocket connection between frontend and backend also allows for streaming responses and intermediate steps to the UI. The different routes provided by the backend can be used to send and receive an automated answer, change the method configuration, get information regarding the agent actions and more. The backend also includes multiple prompting methods that implement different approaches to handle user requests. All methods can then use models either hosted locally via vLLM or LiteLLM, or use externally hosted models, e.g., by OpenAI. Figure~\ref{fig:opaca-llm-arch} shows the overall architecture of SAGE and its peripheral systems. A screenshot of the SAGE frontend can be seen in Figure~\ref{fig:opaca-llm-ui}.

\subsection{Multi-Modality}

Besides basic text input and output, the frontend also provides interaction via voice, using either the browser's built-in TTS/STT support or an external voice server. Further, the UI allows one to select one or more files, which are then sent to the backend and taken into account for the next LLM request, provided that the selected model supports those files, e.g. PDF. The LLM can then summarize those files or use information found in the file as input for later tool calls. External information sources can be included via respective OPACA actions: at the moment, services for Google, Wikipedia, as well as e-mail, weather, and navigation APIs have been implemented and made available to the LLM. The LLM's response is rendered as Markdown, so that e.g. images or tables can be embedded directly into the chat output, and further information can be referred to via hyperlinks.

\subsection{Security and Session Handling}

\Figure()[width=\figwidth]{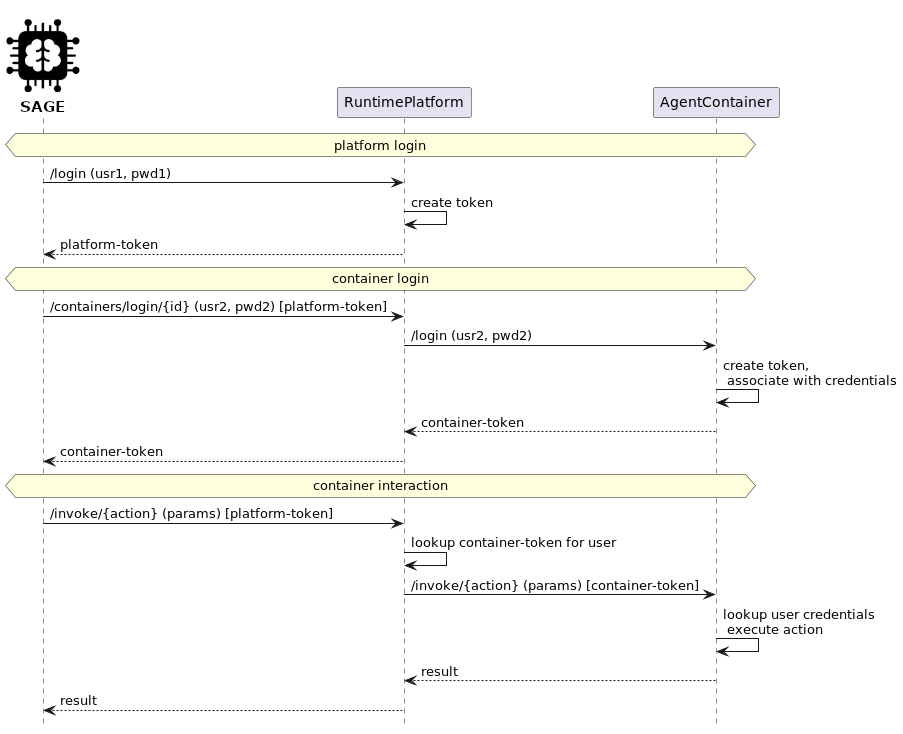}{Authentication to Runtime Platform and Agent Container \label{fig:container-login}}

The OPACA Framework provides basic authentication and user management functions, which are also supported by the SAGE UI by asking for matching credentials when connecting to a platform. The message exchange for the authentication process between SAGE, the OPACA RP, and the AC is displayed in Figure~\ref{fig:container-login}. SAGE further keeps track of multiple sessions, allowing individual users to connect to different runtime platforms, using different configurations, and having individual chat histories.

Additionally, ACs can define their own authentication routes, aligned with the OPACA framework specification for container logins. In the SAGE UI, any service requiring authentication will prompt a secure login to the specific container. Once logged in, the AC can associate those credentials (or an accordingly configured client object) with the current user and automatically retrieve them for each of their future requests.


\section{Different Methods for LLM Prompting}
\label{sec:methods}

SAGE employs different prompt processing strategies, each focusing on different usability aspects such as correctness, answer quality, and speed. In each method, one or multiple ``LLM Modules'' can be used for the final response generation process. 
These modules use the OpenAI API, which has become a de-facto standard and can be used with the actual OpenAI endpoints or self-hosted open source LLM, using e.g. vLLM or LiteLLM as proxy.
Within each method, multiple iterations might occur, prompting each LLM module anew with gained information from the previous iteration. This allows each method to follow a multi-stage solution process as well as an evaluation procedure. All methods interact with the OPACA platform for action retrieval and invocation. 

\begin{itemize}
    \item \emph{Simple/Simple-Tools}: A single LLM module handling all tasks of the tool generation process.
    \item \emph{Tool-Chain}: Two LLM modules splitting the generation and evaluation part.
    \item \emph{Orchestration}: A multi-stage process using a divide \& conquer approach, to divide the given request by the user into smaller subtasks based on the agent composition of the connected OPACA RP.
\end{itemize}

There has been another implementation based on the architecture laid out by Song et al.~\cite{song2023restgpt}, called ``Rest-GPT'', in which four LLM modules were used for selecting, formulating, calling, and evaluating tool calls, respectively. Like the \textit{Simple} method, it was not using the built-in tool calling feature of modern models, relying solely on a combination of thinking- and structured outputs. It has since been deemed outdated since the currently implemented methods surpass its performance while at the same time being easier to maintain.

\Figure()[width=\figwidth]{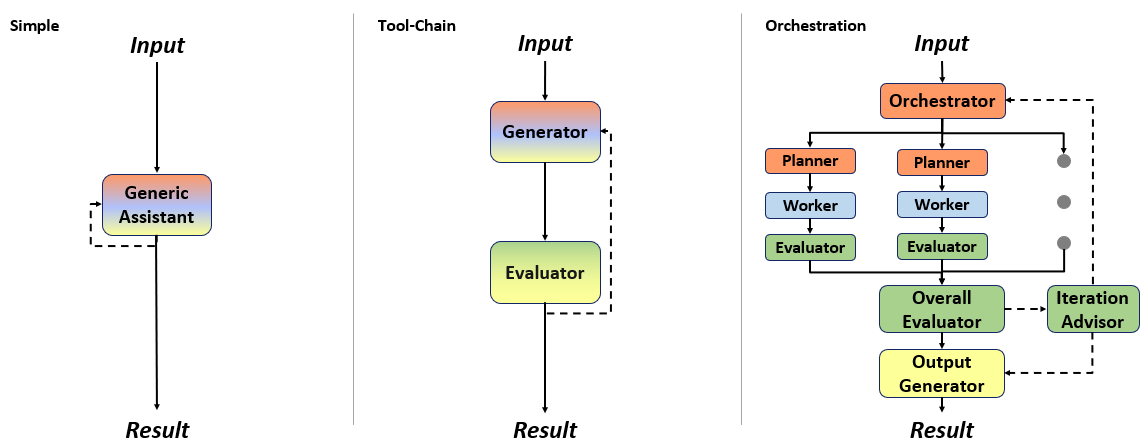}{Overview of the Method Architectures implementing the different modules of the Tool Generation Process in \ref{sec:toolgenprocess} in their LLM modules. (Red: \textit{Planning}, Blue: \textit{Construction}, Green: \textit{Evaluation}, Yellow: \textit{Output Generation}.)\label{fig:method-archs}}

Figure~\ref{fig:method-archs} shows an overview of the LLM modules in the different methods and how they interact with each other. In addition, Table~\ref{tab:methods} provides a comparison of the most relevant features of the different methods.

\begin{table*}[t]
    \centering
    \begin{tabularx}{\textwidth}{@{}XXXXX@{}}
        \toprule
        \bf{Method}  &\bf{Format}&\bf{Parallel}&\bf{LLM Modules}&\bf{LLM Calls}\\
        \midrule
        Simple       & JSON      & no          & $1$           & $n+1$        \\
        Simple-Tools & Tools     & no          & $1$           & $n+1$        \\
        Tool-Chain   & Tools     & yes         & $2$           & $2n$         \\
        Orchestrated & Tools     & yes         & $3k+4$        & $3n+3$       \\
        \bottomrule
    \end{tabularx}
    \caption{Comparison of the tool calling behavior of the different prompting methods: What format tools are provided in, tools can be executed in parallel, how many LLM modules exist, and how many LLM calls are executed. The parameter $k$ stands for the number of agents on the OPACA Platform, and $n$ for the number of (sequential) tools to be called.}
    \label{tab:methods}
\end{table*}

\subsection{Simple}
\label{sec:simple}

The first method that was implemented utilizes a single LLM module, fulfilling all the roles, from tool generation, over tool call evaluation, up to generating the final output. This module is called repeatedly in a loop, feeding it back its previous responses along with the original user query, until the request has been fulfilled or the LLM decides that it cannot be fulfilled.

For this, the \emph{Simple} method relies solely on the normal input and output generation of an LLM: Available tools are injected directly into its system prompt as a list of JSON objects, and likewise the LLM is instructed to output necessary tool calls one at a time in a specific JSON format in its regular chat response content. It is checked whether the generated output conforms to the expected JSON schema of a tool call, and if so, the corresponding OPACA action is invoked with the specified parameters, and the result (or error) is appended to the messages for the next iteration of the loop. This is repeated until the response does not contain a valid tool call; in this case, the response is assumed to be the LLM's final answer and sent back to the user.

The \emph{Simple} method is the only method in our system that does not use the function calling ability of modern models. On the one hand, this approach can support any model, since only the basic query and response will be used. However, relying on interpreting the response as JSON can lead to premature cancellation of the process (outputting the wrongly formatted JSON to the user), especially if the model tends to include ``chatter'' along with the next action to take. In addition, the method is limited to performing a single tool call per iteration.

Despite its simplicity, this approach works well, especially when used with strong models such as \texttt{gpt-4o} or \texttt{gpt-4o-mini}. The general response quality of this approach remains simple, with occasional outputs of attempted action calls in malformed JSON formats. Still, in most situations, the LLM is able to infer the right actions to take, with the right parameters, and also correctly converts the output of one action into the input for the next (e.g., from a tuple of longitude and latitude to an object with those attributes).

\subsection{Simple-Tools}
\label{sec:simple-tools}

The \emph{Simple-Tools} method is a variant of the \emph{Simple} method: It also uses only a single LLM module for all the roles, called in a loop, but in this approach, the integrated ``tools'' feature of modern models is used to pass available tools to the LLM and generate tool calls for user queries within a special field of the LLM's response, performing a structural validation of the tool call beforehand. 

For this, actions acquired from OPACA are converted into a well-defined format, understandable for most LLMs, including their name and description, a list of parameters, and definitions of expected parameter types, similar to JSON Schema. This can easily be generated from OPACA's OpenAPI specification.

As in the \emph{Simple} method, if a tool call is found in the response, the corresponding OPACA action is invoked and the result (or error) is added to the messages for the next iteration of the loop, until the response does not include any more tool calls.

Unlike the \emph{Simple} method, multiple tool calls can be generated and executed in a single iteration of the loop, potentially reducing the number of necessary LLM calls. Also, by making use of the native tools feature, the system prompt can be significantly more simple, while at the same time achieving a more reliable tool generation. The downside, of course, is that it only works with models that support tools, which nowadays applies to most, but not all, LLMs. Further, the maximum number of tools can be limited.

\subsection{Tool-Chain}
\label{sec:tool}

Like \emph{Simple-Tools}, the \emph{Tool-Chain} method makes use of the built-in ``tools'' feature of modern LLMs, but there are two different modules in this method: The first module, the  ``Generator'', will choose and formulate tool calls based on the user input and currently available information, while the second module, the "Evaluator", will summarize the result of these tool calls, after they have been invoked via the OPACA API. Additionally, the Evaluator will determine whether the achieved results are sufficient to deem the user query as fulfilled. If so, the output of the second module will be sent back to the user. Otherwise, the first module will receive the summary from the second module, continuing to output tool calls to fulfill the user request. In the case where no tools are necessary, for example, if the user asks about specific functionalities the LLM provides, the Generator will not generate any tool calls and instead its answer will be directly forwarded to the user.

This method does have limitations, in parts defined by the used model. With a very large multi-agent environment, providing numerous services, the Generator module can get overloaded by the amount of available tools. The current maximum number of tools supported by OpenAI models is 128, which limits the amount of serviceable OPACA actions when using this model family and is the reason, why any additional tools are currently cutoff. Further, a larger amount of available tools will also have an effect on the output quality of the Generator, since a large amount of input tokens negatively affect the performance of an LLM~\cite{liu2023lost}.

The overall response quality is very good, especially since the Evaluator is tasked directly with the summarization of the internal chain, making this method comprehensible and transparent for users. Another special behavior of the Evaluator is the missing system prompt of this module and instead prompting solely with a user message, including the evaluation and summarization task as well as all the tool calls made during this request. This has led to an increase in performance for this module, as opposed to describing the task in the system prompt. However, with two LLM modules this method may increase the necessary time and token usage for an average call, especially when compared against the simple method described previously. Due to the usage of the integrated tool features, the Tool-Chain approach does provide the possibility for parallel tool calls, eliminating the necessity for multiple iterations, and thereby reducing the time for complex user requests involving multiple, parallel tool calls.

\subsection{Orchestration}
\label{sec:orchestration}

As the number of tools grows, smaller open-source models can struggle with function selection, while other models have a hard limit on the number of tools (e.g. 128 for OpenAI) they can handle. To address this challenge, the Orchestration method defines a multi-stage approach, in which user requests are divided into smaller subtasks, based on the agents deployed on the connected RP. The initial user request will be passed to the first LLM module called ``Orchestrator''. This module will have knowledge about all available agents on the connected RP including their actions. However, it does not have detailed information about actions, such as the description and parameters, which will reduce the overall overhead of the initial input. This will also emphasize the importance of agent and action names, since the Orchestrator has only those two information available for a further task breakdown.

As seen in Figure~\ref{fig:method-archs}, the Orchestrator then sends its generated subtasks to one or more so-called ``Agent-Trios'', which mirror the available agents on the connected RP and are therefore initialized dynamically. Each Agent-Trio consists of three LLM modules:

\begin{enumerate}
    \item \textbf{Planner}: Selects the next tool(s) to call
    \item \textbf{Worker}: Formulates the specific tool calls
    \item \textbf{Evaluator}: Evaluates the results sent back by the OPACA platform
\end{enumerate}

The overall functionality of it can be compared to the Tool-Chain approach from the previous section, while further splitting the tool generation part into a Planner and Worker module. The Planner receives a subtask from the Orchestrator, which potentially still needs further breakdown into concrete tool calls. The Worker then receives the selected tool choices by the Planner in natural language and generates concrete tool calls from them, using the tools output field. Afterwards, the corresponding OPACA actions are invoked. The results are evaluated by the Evaluator against the initial task sent by the Orchestrator to this Agent-Trio; if the acquired results satisfy the request, the Agent-Trio is marked as \textit{finished}. 

Once all activated Agent-Trios are \textit{finished}, their results are sent to the ``Overall Evaluator'', which will compare the combined results against the initial user request and will only output its reiteration decision. This is done by providing the model with a guided choice between the output values \textit{REITERATE} and \textit{FINISHED}. In the case of reiteration, the module ``Iteration Advisor'' generates a detailed summary of the last iteration, the called tools and their results, a list of issues, and suggested improvements. It can further decline the request of a reiteration, if it finds its generated output to be sufficient enough to answer the user. The reiteration decision of the Iteration Advisor will be output lastly, using the autoregressive behavior of LLMs~\cite{vaswani2017attention} to make its decision more robust. Finally, all information is passed to the Output Generator, which will formulate the user response. 

This method is significantly more complex than the other methods, which can be both a strength and a weakness. By subdividing the tools across different Agent-Trios, each of those calls can be faster and more reliable, but the orchestration can have problems if different Agent-Trios have similar capabilities, or if tasks require much interaction between different Agent-Trios. The use of a dedicated output generator module requires an additional LLM call, but allows for streaming of the final response, since it splits the evaluation and response generation step.

Another key feature of this architecture is its ability to employ different LLMs for different pipeline stages. In our implementation, we achieved comparable performance to GPT-4o using a combination of \textit{Mistral-Small-24B-Instruct-2501} (for output generation and worker) and \textit{Qwen2.5-32B-Instruct-GPTQ-Int4} (for planning and evaluation tasks). This model configuration will be called ``open-source-composition'' in the following sections. In particular, our experiments revealed that while model quantization generally preserves most capabilities, as suggested in literature~\cite{Huang2024quantization}, it significantly affects the accuracy of tool call generation. This makes quantized models unsuitable for worker roles despite their viability in other pipeline stages. In our tests, we used a 4-bit GPTQ \cite{frantar2022gptq} quantization for all models except a full 16-bit LLM for the Worker module.


\section{Evaluation}
\label{sec:eval}

Performing an in-depth analysis of our constructed system represents a major challenge for a comparable evaluation method. This is due to the unique integration of the OPACA framework and the interaction with tools in the form of OPACA actions. This requires a testing environment to be available as OPACA agent containers. Unfortunately, the availability of public ACs is very limited and existing containers are not meant for evaluation purposes, including actions with varying execution times. Hence, we have constructed specifically designed ACs, implementing services that simulate real-world scenarios. Further, extensive benchmark prompts have been developed, accommodating the available functions by making concrete inputs to our system, expecting a pre-defined result.

\subsection{Data Preparation}

In total, 3 ACs have been constructed, holding 15 agents with a total of 102 actions, providing functionalities in the field of \textit{smart-office}, \textit{warehouse-management}, and \textit{music-player/playlist-management}. The containers have no real-world effect, but maintain state changes, such as the addition or removal of customer orders from the \textit{warehouse-management} container. The complexity of the agents range from simple getter-functions to complex object creations with custom data types. A more detailed description of the containers, agents, and actions, can be found in the public GitHub repository containing the benchmark agents.\footnote{\url{https://github.com/RobertStrehlow/opaca-llm-benchmark-containers}}

The benchmark prompts have all been manually curated and annotated. Overall, we split the prompts into a \textit{Single-Tool} and \textit{Multi-Tool} scenario, which indicates the amount of necessary tool calls a prompt will result in. Prompts from the \textit{Single-Tool} set will only require one tool to be called, while the \textit{Multi-Tool} set will always require at least two tool calls to be made, either being executed in parallel or with information dependencies gained from sequential tool calls. All prompts have been further extended with an expected answer and expected tool calls. The expected answer will be used as a reference for the evaluation with a judge-LLM. The expected tools follow a strict format and include parameter types and values for a more definite comparison of the internal execution chain of SAGE and the expected behavior.

\subsection{Metrics}

During a benchmark test, several metrics are captured to evaluate the overall performance of our system and our implemented methods. Each request captures information about tool calls, token usage, time per LLM module, and how many internal iterations were necessary to formulate the final result. These results are then consolidated and form the overall metrics for a benchmark run of a certain prompting method. An overview of these summarized metrics is given in the following:

\begin{itemize}
    \item \textbf{Response Score}: Value between 1 and 5. Average score of each request during the benchmark run. The score is given by the Judge-LLM by evaluating the expected answer (curated by a human) against the actual answer of SAGE. Does not consider the called tools. 
    \item \textbf{Correct Tool Usage}: Number of requests that have called all expected tools.
    \item \textbf{Perfect Tool Usage}: Number of requests that have called all expected tools \textbf{and} no additional tools.
    \item \textbf{Time}: Total time of the benchmark run.
    \item \textbf{Module Time}: A breakdown of the time each LLM module has spent as part of the answer generation process for each request.
    \item \textbf{Token Usage}: Total number of tokens used.
\end{itemize}

To extract these metrics from our benchmark tests, the following components were implemented:

\subsubsection{Response Score: Judge-LLM}

Each request includes an expected answer, indicating what information or contents should be present in the answer. This allows for a broader range of accepted answers, differing in style and expression. For this, a judge-LLM is tasked to compare the expected answer to the actual output of each request. The judge-LLM is provided with the original request for context and instructions to rate answers on a 1-5 scoring system, as well as to provide a reason for its decision, which can help to manually validate its decision afterwards, if necessary. As a judge, we will use the \textit{gpt-4.1-2025-04-14} model snapshot by OpenAI. For details on how the Judge LLM is prompted and the provided descriptions of each grading scale, please refer to the test code in the public repository.\footnote{\url{https://github.com/GT-ARC/opaca-llm-ui/blob/main/tests/test.py}}

\subsubsection{Tool Checking}

Each request in both benchmark sets (\textit{Single-Tool} \& \textit{Multi-Tool}) specifies their expected tools. This is done by manually curating the name and expected parameters for each tool in each test case. However, these two values were not enough for a proper evaluation, since more sophisticated requests required a better distinction for certain situations. Therefore, the following attributes were further included in our tool checking:

\begin{enumerate}
    \item \textbf{String Strictness}: Some of our simulated services would give specific ids pertaining to a specific name or composition of names. To mimic the correct behavior of these implemented tools, string matching can be set to either \textit{EXACT}, \textit{PARTIAL}, or \textit{NONE}. If an expected tool call would then define a \textit{PARTIAL} string parameter as ``server'', an actual tool call with the used parameter ``Server Room'' would still be labeled as correct.
    \item \textbf{Dependencies}: Regarding the \textit{Multi-Tool} benchmark set, some tools are dependent on previous tools and should only be labeled as correct if the previous call has been made. This should avoid situations in which the request was formulated as ``Do A and then B'', only for the internal chain to execute B right away with correctly guessed parameters.
    \item \textbf{Alternatives}: Certain tool calls will provide the same information. Since our system is primarily focused on fulfilling requests for a user, we allow for alternative calls for specific tools. The metric \textit{Perfect Tool Usage} will then be considered valid, if only one of the grouped alternatives has been called.
    \item \textbf{Optionality}: Some tools or parameters can be optionally included, but their presence or absence should not alter the validation of a tool call if they are not influencing the overall result.
\end{enumerate}

By applying these attributes to the tool checking part in the benchmark script, the validation of correct tool calls is further strengthened by additional insights. This is important to showcase an additional metric, besides the response score given by the Judge-LLM, which is not fully deterministic and may include a certain bias in its automatic evaluation~\cite{ye2024justiceprejudicequantifyingbiases}.

\subsubsection{Resource Monitoring}

SAGE automatically keeps track of the number of internal iterations, completion- and response-tokens and execution time for each request and stores them along the final response. It also breaks down the time it took each module to output text or a tool call. The LLM module of a specific method time might not add up to the overall time, since the tool calling step done by the system itself is not included in those times. For the methods \textit{Simple}/\textit{Simple-Tools}, such a breakdown is unnecessary, since only one LLM module is used. These metrics are then extracted and aggregated by the benchmark script.

\subsection{Benchmark Results}

\begin{figure*}[!t]
    \centering
    \captionsetup[subfloat]{justification=raggedright}
    \subfloat[Correct and Perfect Tool Calls percentage achieved during the Single- and Multi-Tools benchmark runs. All LLM modules in all methods were using the \textit{gpt-4o-mini-2024-07-18} model.]{
        \includegraphics[width=0.47\textwidth]{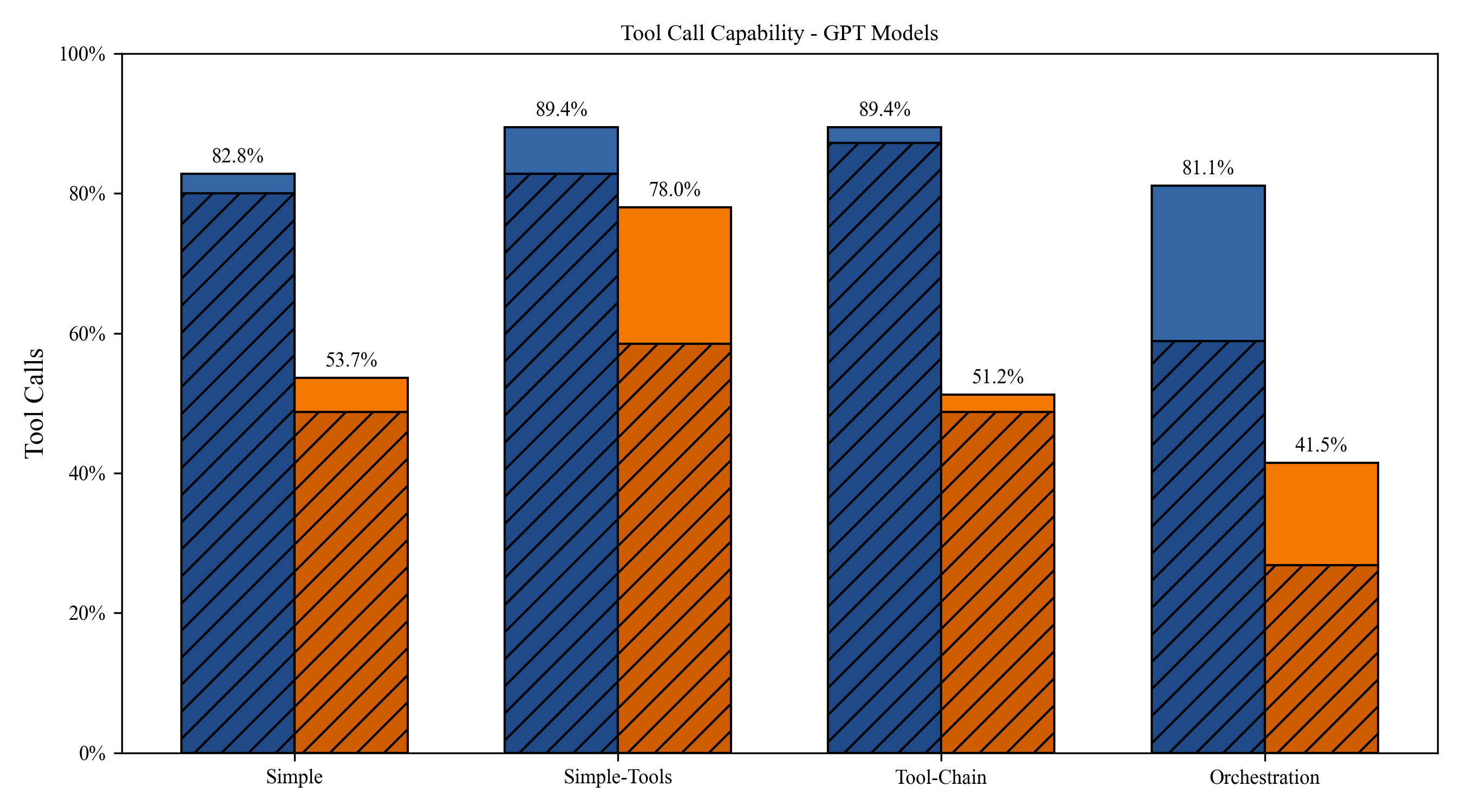}
        \label{fig:simple-tool-usage}
    }
    \hfill
    \subfloat[Correct and Perfect Tool Calls percentage achieved during the Single- and Multi-Tools benchmark runs. The \textit{Orchestration} method uses the OSC (see Section~\ref{sec:orchestration}). All remaining methods used only the \textit{Mistral-Small-24B-Instruct-2501} model for their LLM modules.]{
        \includegraphics[width=0.47\textwidth]{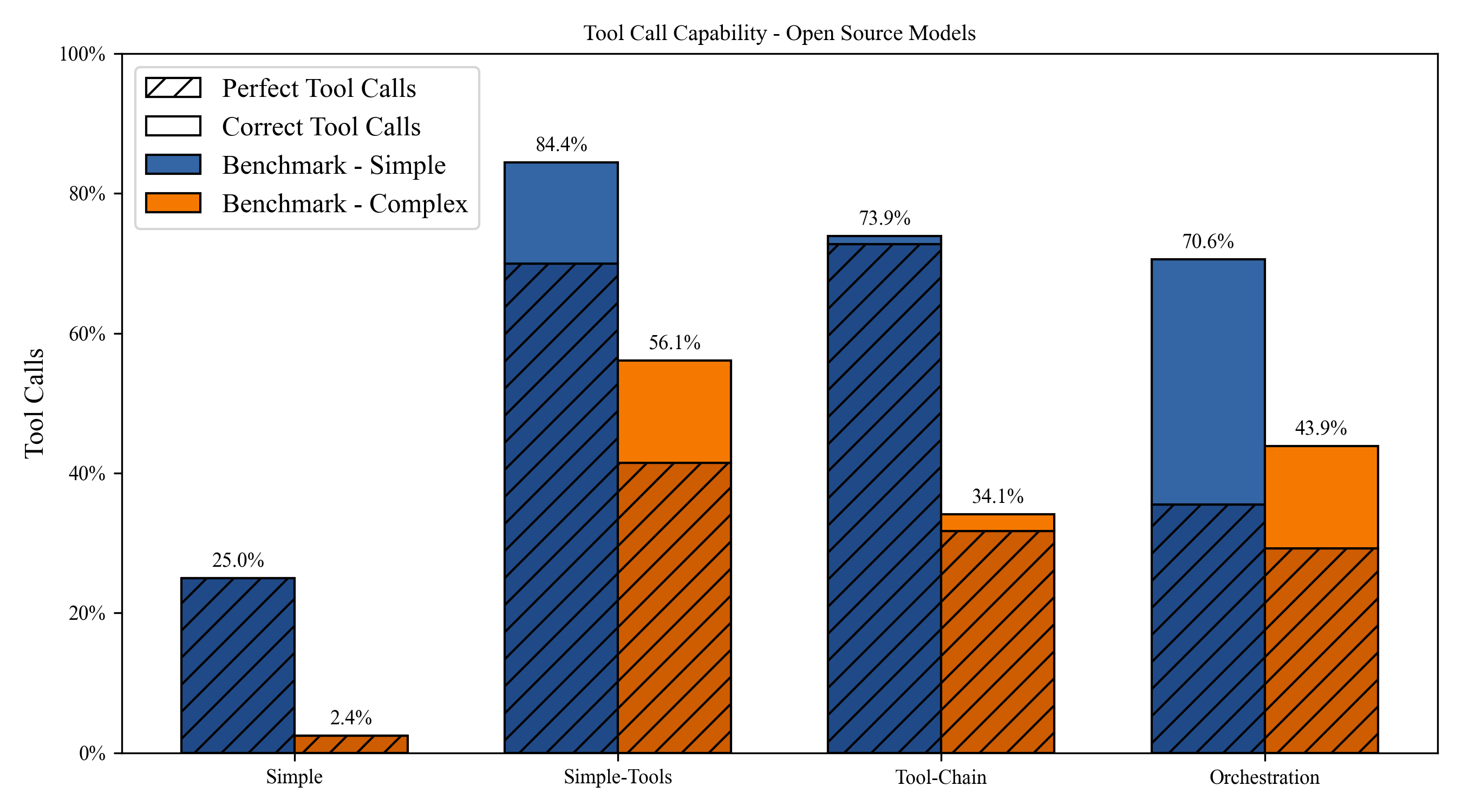}
        \label{fig:complex-tool-usage}
    }
    \caption{Comparison of tool usage across different benchmark types.}
    \label{fig:tool-usage-all}
\end{figure*}

We have conducted multiple tests, using the aforementioned ACs and benchmark prompts. In this section, we will present the gathered results, comparing our implemented methods and showcase the effectiveness of our constructed system. Further, a discussion and evaluation of the achieved results will be made, reflecting on the set goals and translation ability of our system into other scenarios. The complete results files are available on the project's GitHub page\footnote{\url{https://github.com/GT-ARC/opaca-llm-ui/tree/main/tests/benchmark_results}}.

\subsubsection{Final Response Score and Efficiency Metrics}

For better comparability, the results in Table~\ref{tab:score-results} are taken from tests using the same model across all methods. The Judge-LLM uses the same model for all comparisons (\textit{gpt-4.1-2025-04-14}). Every given score for a single generated output of a request also includes a reason, making the assessment of the Judge-LLM understandable for a human. The scores range from 1.0, being the worst possible score, to 5.0, being the best possible score. 

The remaining metrics in Table~\ref{tab:score-results}, \textbf{Time} and total \textbf{Token} usage, are accumulated for each prompt given to the system using the specified method. This is possible since the benchmark prompts are the same for every method test.

\begin{table}[t]
    \centering
    \footnotesize  
    \begin{tabularx}{\columnwidth}{ 
        X  
        S@{\,/\,}S
        S@{\,/\,}S
        X 
    }
        \toprule
        \textbf{Method} & \multicolumn{2}{c}{\textbf{Score (1–5)}} &
        \multicolumn{2}{c}{\textbf{Time (s)}} &
        \textbf{Tokens (Mio.)} \\
        \midrule
        Simple          & 4.52 & 3.81 & {\bfseries 2.36} & {\bfseries 6.14} & 3.478 
        \vspace{0.2em} \\
        Simple-Tools    & 4.68 & {\bfseries 4.20} & 5.24 & 8.45 & 3.120 
        \vspace{0.2em} \\
        Tool-Chain      & {\bfseries 4.72} & 4.00 & 6.81 & 9.99 & 3.036 
        \vspace{0.2em} \\
        Orches-tration   & 4.16 & 3.71 & 14.26 & 16.98 & {\bfseries 2.557} \\
        \bottomrule
    \end{tabularx}
    \caption{Method comparison}
    \label{tab:score-results}
\end{table}

Each method exceeds in a different area for the three metrics. While the \textit{Simple} method uses the most tokens, it also takes the least amount of time by a large margin. The \textit{Simple-Tools} and \textit{Tool-Chain} methods are relatively close in all areas, but will show distinctive characteristics in the tool usage comparison. The \textit{Orchestration} method performs the worst while taking the longest as well, however, its token usage is the lowest overall.

With the method's architectures in mind, these results were expected to an extend. The \textit{Simple} method only uses a single LLM modules while also generating function calls in a JSON schema in its normal output, resulting subsequently in a very low time for the benchmark test. The high number of tokens can also be explained by the fact, that the \textit{Simple} method is the only method unable to formulate multiple function calls in a single LLM inference. For this reason, it is further the only method to not set a limit for internal iterations, only stopping if no valid JSON format was being output, as described in Section~\ref{sec:simple}. 

The \textit{Simple-Tools} and \textit{Tool-Chain} methods are very comparable regarding the metrics in Table~\ref{tab:score-results}. Their inner structure only differs from an additional LLM-component in the \textit{Tool-Chain} method, which is encapsulated in the single LLM modules of the \textit{Simple-Tools} method.

The \textit{Orchestration} is a special case, since it raises the minimum used LLM modules to six per request, but performs overall the worst and slowest of all methods. However, its redeeming quality is the low number of used tokens, since it follows a hierarchical approach of available functions, in which each function is only accessible by its agent, which itself is only provided to the orchestrator module with a general description. The strategy closely resembles a divide \& conquer approach, minimizing especially the input tokens for an LLM-call.

\subsubsection{Tool Selection}

We now compare the valid tool calling ability of our methods and further compare this ability between our proprietary used model (\textit{gpt-4o-mini-2024-07-18}) and the open source models described especially in Section~\ref{sec:orchestration}. The Mistral and Qwen model series were chosen in particular, due to their wide range of available models and comparability to the GPT series of OpenAI.

Unlike the Judge-LLM response score, both the \textit{Simple-Tools} and \textit{Tool-Chain} excel in their tool formulating capability in both prompting scenarios against the other methods. The most noticeable difference between those two methods is the ability of the \textit{Tool-Chain} method to formulate the correct tool call on the first attempt more often than the \textit{Simple-Tool} method, resulting in more relative \textit{Perfect Tool Usage} results.

The \textit{Simple} method especially struggles with the open-source model, most likely due to the custom schema that is necessary to lead to correct tool calls. This theory is supported by the correct tool formulation in other methods, which use the ``tool\_call'' output field of an LLM inference. However, the \textit{Simple} is very strongly influenced by the used model, since models that tend to output a lot of ``chatter'' alongside its JSON schema output will also lead to a preemptive cancellation of the internal chain, before the final answer could be reached.

The \textit{Orchestration} method present the steepest relative drop from \textit{Correct Tool Usage} to \textit{Perfect Tool Usage}, indicating a lot of unnecessary tool call formulation during the response generation. However, it is also the only method that has better or equal results while using open-source models, although only in the \textit{Multi-Tool} scenario by a very small margin.

\subsubsection{Resource Efficiency}

\begin{figure}[!t]
    \centering
    \captionsetup[subfloat]{justification=centering}
    \subfloat[gpt-4o-mini-2024-07-18]{
        \includegraphics[width=0.47\columnwidth]{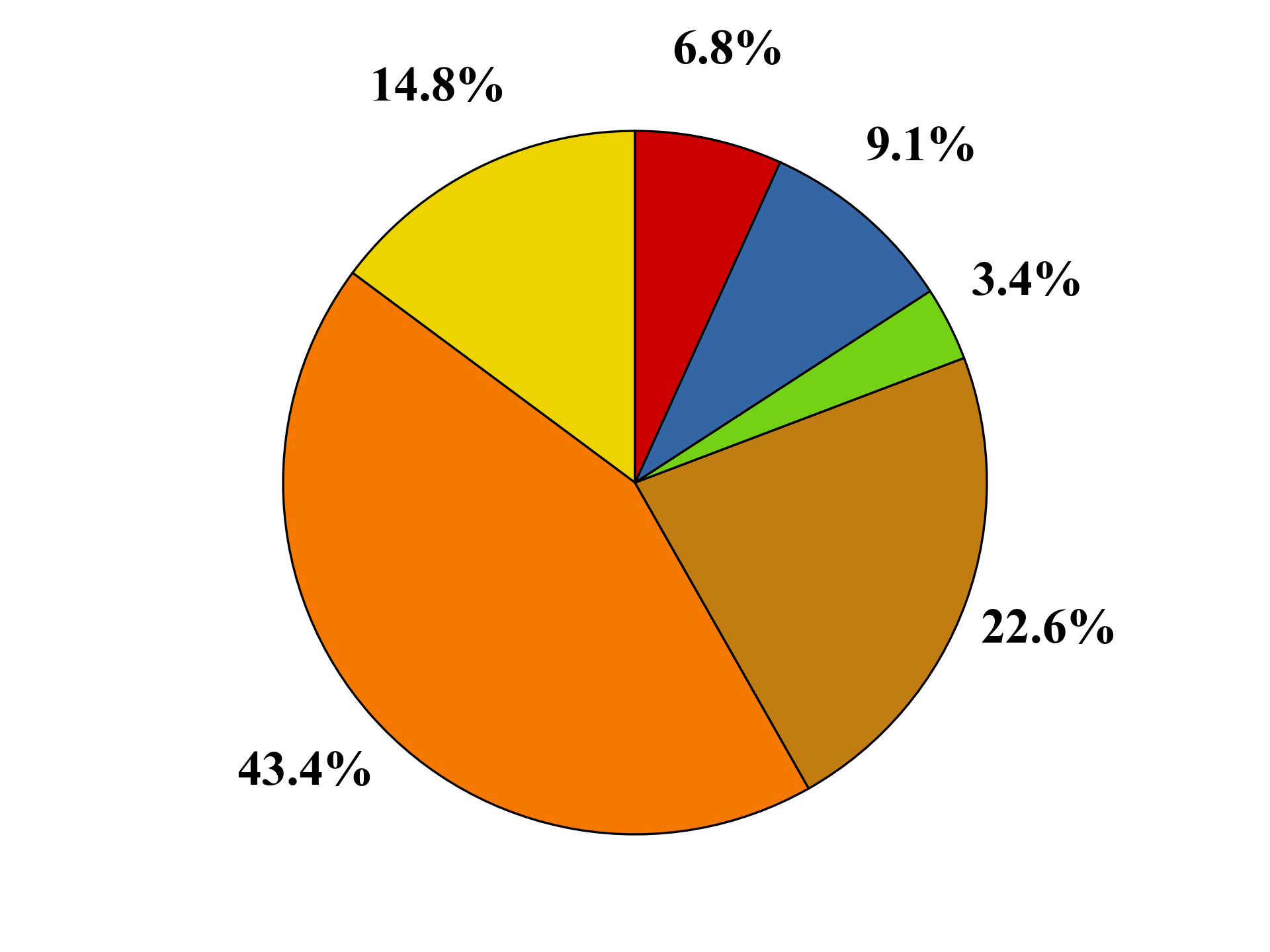}
    }
    \hfill
    \subfloat[Open-Source Composition]{
        \includegraphics[width=0.47\columnwidth]{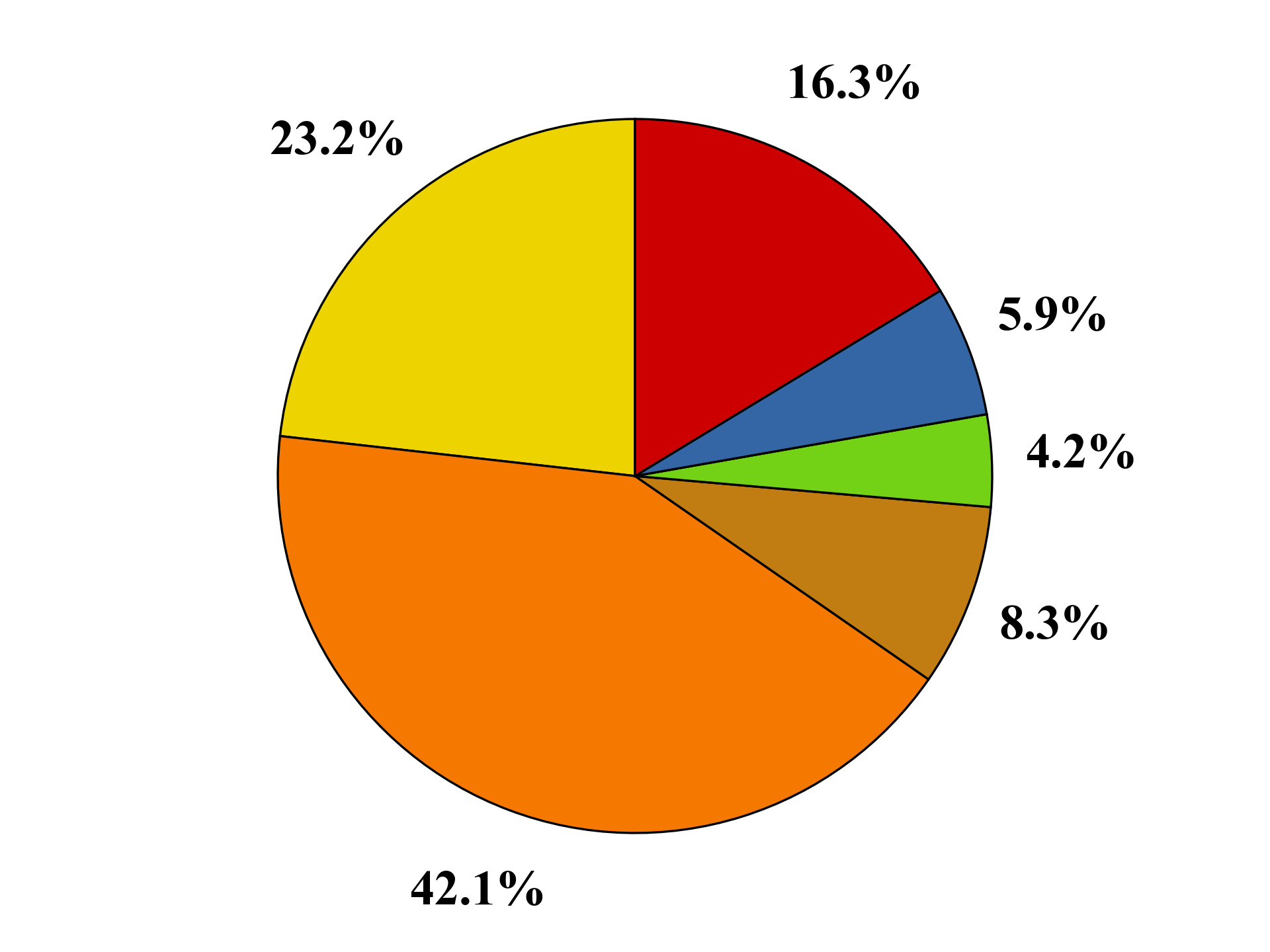}
    }
    \caption[caption]{LLM modules time distribution of the \textit{Orchestration} method with two models for each benchmark prompting set. The colors correspond to the following LLM modules (Yellow: \textit{Orchestrator}, Orange: \textit{Planner}, Brown: \textit{Worker}, Green: \textit{OverallEvaluator}, Blue: \textit{OutputGenerator}, Red: \textit{IterationAdvisor}).}
    \label{fig:comp_times}
\end{figure}

The ability to use given resources efficiently, especially in regard to computational needs, is crucial for an effective tool call generation strategy. To measure these metrics, each benchmark test additionally captures the execution time provided by each implemented model itself and additionally provides the distribution of the time for each LLM module within the selected method. For the \textit{Simple}/\textit{Simple-Tools} method, this distribution is not given, since the total time corresponds to the only LLM module used within these methods. The distribution for the \textit{Tool-Chain} method is for all configurations very balanced between the \textit{Generator} and \textit{Evaluator} modules, with a roughly 50\%-50\% distribution between them. However, a larger variance can be observed for the \textit{Orchestration} method, showing a clear difference between the two tested model configurations as well as the two different prompting scenarios.

In Figure~\ref{fig:comp_times}, the results of the two benchmarking prompts have been averaged. For all benchmark tests of the \textit{Orchestration} method, the \textit{Worker} is playing a crucial role in the tool call generation process, by deciding what tools need to be called next for the given subtask, showcased by its disproportional portion size in the overall time consumption of this method. Comparing the two used models with their previous tool call construction rate in mind, the open-source composition is using the \textit{Orchestrator} as well as the \textit{IterationAdvsisor} more dominant. This is most likely due to the amount of corrections the internal chain of LLM modules have to apply, resulting in a new invocation of the \textit{Orchestrator}, which, in an ideal tool planning phase, should have only been invoked once. The \textit{IterationAdvisor} is also only called if the \textit{OverallEvaluator} has decided that a reiteration is necessary. Clearly, the open-source composition struggles with finding the right tool calls on the first attempt, which is further showcased in Figure~\ref{fig:tool-usage-all} indicating a large portion of tool calls were not ``perfect'', but rather just correct after numerous attempts of the internal chain.

Overall, the implemented methods have shown the reliable tool calling performance of the system, with different strengths regarding the presented metrics. Consequently, each method has a different use-case in which it can outperform the other methods, although for general tasks, the \textit{Simple-Tools} and \textit{Tool-Chain} method could be preferred, since their overall tool calling ability is a good choice for basic tasks with a relatively fast response time.


\section{Conclusion}
\label{sec:conclusion}

In this work, we have presented, analyzed, and proved the effectiveness of our constructed LLM-based methods in zero-shot inferences, formulating tool calls provided by the OPACA framework as agent actions and calling them over the OPACA RP. Our system ``SAGE'' is further extendable, making it a solid choice for scalable environments, while also introducing authentication and user-management to the connected multi-agent-system and individual agent container. The adoption of the OPACA framework in an already existing software environment is minimally invasive, while also providing all advantages of an agentic microservice architecture.

The implemented methods have demonstrated different strengths in various categories, important for the inference of LLMs, especially while formulating tool calls, such as valid tool formulation, response time, token usage, or an automated response score given by a judge-LLM. The ability to use multiple models, proprietary and open-source alike, is another advantage of our implemented system, making the adaption of newly released models, such as GPT-5, easy and without much effort.

However, even with more advanced models, intelligent, multi-stage knowledge-flows have a huge potential in complex service environments. Especially the inclusion of OPACA's abstraction layers of Agent Containers, Agents, and Actions can help to reduce the overhead produced by available tool call definitions. This way, a larger amount of tool calls can be supported by a single LLM-system without reducing the overall performance, by leveraging core concepts of multi-agent technologies.

\subsection{Outlook}

Besides constantly working on and improving the different prompting methods, their reliability and performance, we are currently investigating ways to include more `proactive' and `introspective' functionalities into SAGE. Those would, e.g., allow the system not only to adapt and re-configure itself based on the users' instructions or (inferred) preferences, but also to schedule and manage tasks for (regular) execution at a later time. Those and similar additions will enhance SAGE’s capability to support improved productivity in daily work.

\section*{Acknowledgments}

We want to thank the entire development team of SAGE and OPACA, former and current, not all of whom were able to contribute to this article, in particular Cedric Braun, Daniel Wetzel, Tolga Akar, Benjamin Acar, and Abdullah Kiwan.

\bibliographystyle{plain}
\bibliography{literature}

\EOD

\end{document}